\documentclass[doublecol]{epl2} 

\title{Optimal working conditions for thermoelectric generators with\\ realistic thermal coupling}
\shorttitle{Optimal working conditions for thermoelectric generators} 

\author{Y. Apertet\inst{1} \and H. Ouerdane\inst{2} \and O. Glavatskaya\inst{3}\inst{4} \and C. Goupil\inst{4} \and Ph. Lecoeur\inst{1}}
\shortauthor{Y. Apertet \etal}

\institute{                    
  \inst{1} Institut d'Electronique Fondamentale, Universit\'e Paris-Sud CNRS, 91405 Orsay, France\\
  \inst{2} CNRT Mat\'eriaux UMS CNRS 3318, 6 Boulevard Mar\'echal Juin, 14050 Caen Cedex, France\\
  \inst{3} Renault SAS, SAS FR TCR AVA 058, 1 avenue du Golf, 78288 Guyancourt, France\\
  \inst{4} Laboratoire CRISMAT, UMR 6508 CNRS, ENSICAEN et Universit\'e de Caen Basse Normandie, 6 Boulevard Mar\'echal Juin, F-14050 Caen, France
}

\abstract{We study how maximum output power can be obtained from a thermoelectric generator (TEG) with nonideal heat exchangers. We demonstrate with an analytic approach based on a force-flux formalism that the sole improvement of the intrinsic characteristics of thermoelectric modules including the enhancement of the figure of merit is of limited interest: the constraints imposed by the working conditions of the TEG must be considered on the same footing. Introducing an effective thermal conductance we derive the conditions which permit maximization of both efficiency and power production of the TEG dissipatively coupled to heat reservoirs. Thermal impedance matching must be accounted for as well as electrical impedance matching in order to maximize the output power.
Our calculations also show that the thermal impedance does not only depend on the thermal conductivity at zero electrical current: it also depends on the TEG figure of merit. Our analysis thus yields both electrical and thermal conditions permitting optimal use of a thermoelectric generator.}

\begin{document}

\maketitle

\section{Introduction}

Thermoelectric power generation poses challenges which are of fundamental and technological nature. The development of efficient thermoelectric systems is widely recognized as a strategic topic of applied research in  view of problems related to, e.g., waste heat recovery and conversion to electricity. This is reflected by an abundant litterature on the subject~\cite{rowe}. Recent progress in technological development of TEGs has relied on advances in material sciences: new materials and new techniques to produce specific structures have permitted the improvement of device performance through the characterization and optimization of their electrical and thermal transport properties (see the review of Di Salvio \cite{disalvio} and the recent one of Shakouri \cite{shakouri}). The performances still are quite modest though and, as for all heat engines, thermoelectric generators are subjected to the laws of thermodynamics, which impose an upper bound to their efficiency, the so-called Carnot efficiency, $\eta_{\rm C}$. Much effort thus is invested to seek ways to improve the intrinsic properties of thermoelectric modules and hence approach the upper efficiency limit.

These properties often are summarized into the paradigmatic figure of merit $ZT$, which characterizes the performance of a device at average temperature $T$~\cite{rowe} (a precise definition of the quantity $Z$ is given in the next section). As an illustration of progress we mention the recent results of Snyder and coworkers: with p-type doping of PbTe-based semiconductor materials, i.e. through band structure engineering, a value of $ZT$ as high as 1.8 has been reported \cite{snyder1}. This impressive value was obtained at a temperature of 850 K, which means that much progress remains to be done to achieve this level of performance at room temperature. Another aspect of the problem was pointed out by Nemir and Beck\cite{nemirbeck}: since the figure of merit is determined by a set of three parameters (the electrical conductivity, the thermal conductivity, and the thermopower), there is in principle an infinite number of possibilities to obtain a given value of $ZT$, and one direct consequence of this fact is that $ZT$ alone is insufficient to characterize the performance of a thermoelectric system. Furthermore, as discussed below, the ability for a TEG to deliver a high output power also depends on impedance matching which must account for thermal resistance; this point directly relates to the more general problem of the optimization of the working conditions of a non-endoreversible thermodynamic engine under specific constraints, which indeed differ from those of endoreversible engines, also known as the Novikov-Curzon-Ahlborn (NCA) conditions\cite{Chambadal,Novikov,curzahl,bejan96}.

In this work, we thus adopt a particular viewpoint: we do not seek ways to obtain ever higher values of $ZT$; rather, we want to understand how one can optimize both electrical and thermal conditions so that a thermogenerator may produce a maximum ouput power. Indeed, if the connections of a TEG to two heat reservoirs are assumed to be ideal, the best working conditions are well known; but when thermal dissipative couplings exist, as for all real systems, these conditions turn out to be different. A global reflexion on realistic working conditions of TEGs thus is necessary to advantageously exploit the benefits of the works on the materials properties. Much work has already been devoted to the modeling and simulation of thermoelectric devices, but the thermodynamic optimization has been a subject of debate, especially when considering the maximal efficiency versus maximal power strategy \cite{clingman,snyder04,seifert10}.

In our work, we consider that the electrons in the TEG form a carrier gas that can be described in the frame of linear out-of-equilibrium thermodynamics developped by Onsager\cite{onsager1,onsager2} and Callen\cite{callen1,callen2}. This approach is most convenient to study on the same footing the thermodynamic forces inducing the irreversible processes in a system, and the response of this system in terms of fluxes. If the fluctuations are sufficiently small a linear force-flux coupling provides a correct description of a TEG since the physics of such systems is based on the interplay between Ohm's law and Fourier's law. Our article is organized as follows. In section 2, we introduce the model thermoelectric generator, the definitions and notations we use throughout the paper. In Section 3, we analyze the electrical and thermal conditions for output power and efficiency maximization considering non-ideal thermal contacts and a fixed figure of merit. We end the paper with a discussion and concluding remarks.

\section{Model of thermoelectric generator}

\begin{figure}
\scalebox{.42}{\includegraphics*{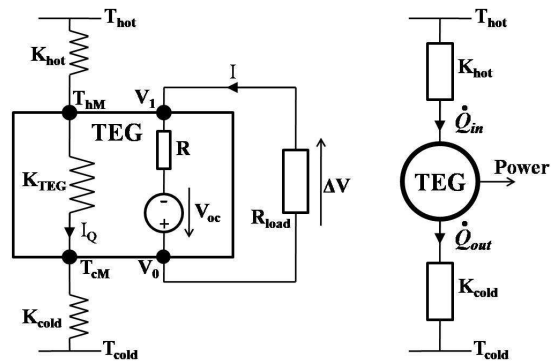}}
\caption{\label{teg} Thermoelectrical (left) and thermodynamical (right) pictures of the thermoelectric module and the load.}
\end{figure}

We consider a single-leg module placed between two ideal heat reservoirs as depicted in Fig.~\ref{teg} (the doping type does not affect the generality of the conclusions of our study). The temperatures of the reservoirs are denoted $T_{\rm cold}$ (colder one) and $T_{\rm hot}$ (hotter one) respectively. The two thermal contacts are characterized by the thermal conductances $K_{\rm cold}$ and $K_{\rm hot}$, respectively. The TEG is characterized by its isothermal electrical resistance $R$, its thermal conductance $K_{\rm TEG}$, and the Seebeck coefficient $\alpha$. The thermal conductance $K_{\rm TEG}$ reduces to the conductance $K_{_{V=0}}$, under zero voltage (electrical short circuit), and to the conductance $K_{_{I=0}}$, at zero electrical current (open circuit). Electrons and phonons contribute to the thermal conductance, and $Z$ is given by : $Z = \alpha^2/RK_{_{I=0}}$.

\subsection{Thermal and electrical currents}

In real systems, the heat exchangers between the temperature reservoirs and the thermoelectric engine contribute to the energy loss and the related decrease of efficiency. In other words, entropy is produced in the engine as well as in the exchangers. The thermodynamic system we consider is more complex than an NCA one, and assuming that the incoming and outgoing heat fluxes are linear in the temperature difference (see Ref.~\cite{huleihil} for further detail on the laws governing heat fluxes), we describe the TEG characteristics with the force-flux formalism, which yields the following equation:
\begin{equation}\label{frcflx}
\left(
\begin{array}{c}
I\\
I_Q\\
\end{array}
\right)
=
\frac{1}{R}~
\left(
\begin{array}{cc}
1~ & ~\alpha\\
\alpha \overline{T}~ & ~\alpha^2 \overline{T} + RK_{_{I=0}}\\
\end{array}
\right)
\left(
\begin{array}{c}
\Delta V\\
\Delta T'\\
\end{array}
\right),
\end{equation}
\noindent where $I$ is the electrical current through the load, and $I_Q$ is the thermal current; $\Delta V$ and $\Delta T' = T_{\rm hM}-T_{\rm cM}$ are the voltage and the temperature difference across the TEG, $T_{\rm hM}$ and $T_{\rm cM}$ being respectively the temperatures on the hot and cold sides of the TEG. The open-circuit voltage is $V_{\rm oc} = \alpha\Delta T'$. The average temperature in the module is taken as $\overline{T} = (T_{\rm cM}+T_{\rm hM})/2$. Since we assume that the system's response is linear, the temperature difference $\Delta T'$ is necessarily small compared to the mean temperature. We show below how an expression for the TEG thermal conductance $K_{\rm TEG}$ can be derived using two different ways.

Within the force-flux formalism the thermal current is expressed as the sum of the contributions of convective heat transfer and steady-state conduction:
\begin{equation}\label{IQ1}
I_Q = \alpha \overline{T} I + K_{_{I=0}} \Delta T'
\end{equation}

\noindent Assuming that the load is simply resistive, Ohm's law applies as follows: $\Delta V = -R_{\rm load} I$ and the electrical current $I$ reads:
\begin{equation}\label{Int1}
I = \frac{\Delta V + \alpha \Delta T'}{R} = \frac{\alpha \Delta T'}{R_{\rm load} + R}
\end{equation}

\noindent This expression of the electrical current $I$ is fed to Eq.~(\ref{IQ1}), which defines the TEG thermal conductance $K_{\rm TEG}$:
\begin{equation}\label{IQ2}
I_Q = \left( \frac{\alpha^2 \overline{T}}{R_{\rm load} + R} +K_{_{I=0}} \right)\Delta T' = K_{\rm TEG}\Delta T'
\end{equation}

\noindent We see that the thermal conductivity $K_{\rm TEG}$ depends on the electrical operating point since its expression given above contains the load electrical resistance.

Now, it is instructive to derive $K_{\rm TEG}$ in a different fashion. The relationship between the two thermal conductances $K_{_{V=0}}$ and $K_{_{I=0}}$ of the TEG \cite{goupil2}:
\begin{equation}\label{thermconds}
K_{_{V=0}} = K_{_{I=0}} \left( 1+ Z\overline{T}\right)
\end{equation}

\noindent can be extended to the following phenomenological formula:
\begin{equation}\label{kteg2}
K_{\rm TEG}(I) = K_{_{I=0}} \left( 1+ \frac{I}{I_{\rm sc}} Z\overline{T}\right),
\end{equation}

\noindent where $I_{\rm sc} = \alpha\Delta T'/R$, is the short circuit current such that: $K_{\rm TEG}(I_{\rm sc}) = K_{_{V=0}}$. It is easy to check that equations (\ref{kteg2}) and (\ref{IQ2}) yield exactly the same expression for $K_{\rm TEG}$. This approach is interesting for two reasons: first, the phenomenological law (\ref{kteg2}) is validated, not solely because the relationship between the two thermal conductances $K_{_{V=0}}$ and $K_{_{I=0}}$ is recovered for $I=I_{\rm sc}$; second, it is important to note that since the short circuit current $I_{\rm sc}$ depends on the effective temperature difference across the TEG, $\Delta T'$, there is no closed form solution for the global currents and potentials distributions in the TEG.

\subsection{Temperatures across the TEG}

The analysis developped so far assumes the knowledge of $\Delta T'$, but it is more useful to obtain expressions of power and efficiency as functions of the temperature difference between the two reservoirs, $\Delta T = T_{\rm hot} - T_{\rm cold}$. Here, we give a brief outline of the calculations that yield the relationships between $T_{\rm cM}$, $T_{\rm hM}$, $T_{\rm cold}$, and $T_{\rm hot}$. First, we define the incoming heat flux: $\dot{Q}_{\rm in} = K_{\rm hot}(T_{\rm hot} - T_{\rm hM})$ and outgoing heat flux: $\dot{Q}_{\rm out} = K_{\rm cold}(T_{\rm cM} - T_{\rm cold})$; following Ioffe's approach \cite{Ioffe}, these may also be written as:
\begin{eqnarray}
\label{Qin}
\dot{Q}_{\rm in} &=& \alpha T_{\rm hM}I - \frac{1}{2} RI^2 + K_{_{I=0}}(T_{\rm hM} - T_{\rm cM})\\
\label{Qout}
\dot{Q}_{\rm out} &=& \alpha T_{\rm cM}I + \frac{1}{2} RI^2 + K_{_{I=0}}(T_{\rm hM} - T_{\rm cM})
\end{eqnarray}

\noindent These equations yield a 2$\times$2 system which links $T_{\rm cM}$ and $T_{\rm hM}$ to $T_{\rm cold}$ and $T_{\rm hot}$:
\begin{equation}\label{temps}
\left(
\begin{array}{c}
\phantom{-}T_{\rm hot~} + \frac{\displaystyle 1}{\displaystyle 2}\frac{\displaystyle RI^2}{\displaystyle K_{\rm hot~}}\\
~\\
-T_{\rm cold} - \frac{\displaystyle 1}{\displaystyle 2}\frac{\displaystyle RI^2}{\displaystyle K_{\rm cold}}\\
\end{array}
\right)
=
\left(
\begin{array}{cc}
{\mathcal M}_{11} & {\mathcal M}_{12}\\
~\\
{\mathcal M}_{21} & {\mathcal M}_{22}\\
\end{array}
\right)
\left(
\begin{array}{c}
T_{\rm hM}\\
~\\
T_{\rm cM}\\
\end{array}
\right),
\end{equation}

\noindent where the four dimensionless matrix elements are given by:
${\mathcal M}_{11} =  K_{_{I=0}}/K_{\rm hot} + \alpha I/K_{\rm hot} +1$, 
${\mathcal M}_{12} =  - K_{_{I=0}}/K_{\rm hot}$, 
${\mathcal M}_{21} =  K_{_{I=0}}/K_{\rm cold}$, 
${\mathcal M}_{22} =  \alpha I/K_{\rm cold} - K_{_{I=0}}/K_{\rm cold} - 1$.

\noindent The analytic expressions of the temperatures $T_{\rm hM}$ and $T_{\rm cM}$ are easily obtained by matrix inversion, but the exact expression of $\Delta T'$ as function of $T_{\rm hot}$ and $T_{\rm cold}$ is cumbersome, and not necessary for the discussion in the subsequent part of the article. Instead, it is worthwhile to seek an approximate but straightforward relationship between $\Delta T'$ and $\Delta T$, which we do as follows. Introducing the total contact conductance,  $K_{\rm contact}$, as $K_{\rm contact}^{-1} = K_{\rm cold}^{-1}+K_{\rm hot}^{-1}$, and assuming that the thermal flux is constant in the whole system, we obtain the following simple relation between $\Delta T'$ and $\Delta T$:
\begin{equation}\label{kapctc}
\Delta T' = T_{\rm hM} - T_{\rm cM} \approx \frac{K_{\rm contact}}{K_{\rm TEG}+K_{\rm contact}}~\Delta T
\end{equation}

\noindent using an analogue of the voltage divider formula. The assumption we just made amounts to consider that the produced electrical power is negligible in comparison to the thermal current entering the thermoelectric generator, and it thus holds well when the temperature difference $\Delta T$ is not too large: in this case the Carnot efficiency is small, and the real efficiency even much smaller. Note that equation~(\ref{kapctc}) still holds for the dissymetric case ($K_{\rm cold}\neq K_{\rm hot}$). In such situation, the dissymetry is contained in $\overline{T}$: the mean temperature gets closer to $T_{\rm hot}$ ($T_{\rm cold}$) if $K_{\rm cold}$ is smaller (greater) than $K_{\rm hot}$. However, these variations are small since $\overline{T}$ is comprized between $T_{\rm cold}$ and $T_{\rm hot}=T_{\rm cold}+\Delta T$: $\overline{T}$ is approximately equal to the mean temperature between heat reservoirs.

\subsection{Th\'evenin generator model}
In Fig.~\ref{teg} the electrical part of the TEG is viewed as the association of a perfect generator and a resistance which is the physical resistance of the generator. The open circuit voltage depends on the temperature difference seen by the TEG: $V_{\rm oc}=\alpha \Delta T'$. In the presence of finite thermal contacts the temperature difference $\Delta T'$ depends on the electrical load, hence the tension generator can no longer be considered as perfect since its characteristics depend on the load. To express this dependence explicitly, we insert the expression of $K_{\rm TEG}$ given by Eq.~(\ref{kteg2}) and the definition of the short circuit current $I_{\rm sc}$ into Eq.~(\ref{kapctc}), and we find $V_{\rm oc}$ as the sum of two terms:
\begin{equation}\label{voc}
V_{\rm oc} =\alpha \Delta T \frac{K_{\rm contact}}{K_{{_{I=0}}}+K_{\rm contact}}-I R \frac{Z\overline{T}}{1+K_{\rm contact}/K_{_{I=0}}}
\end{equation} 

\noindent The first term on the right hand side is independent of the electrical load, the second depends on the electrical current delivered: $V_{\rm oc} = V_{\rm oc}' - IR'$. We thus obtain a rigorous Th\'evenin modeling of the electrical part of the TEG with the definitions of the open circuit voltage given by $V_{\rm oc}'$ and the internal resistance is $R_{\rm TEG}=R+R'$.

\section{Maximization of power and efficiency with nonideal thermal contacts and fixed $Z\overline{T}$}

\subsection{Maximization of power by electrical impedance matching}

The electrical power produced by the TEG can be simply expressed as
\begin{equation}\label{Pprod1}
P = VI = \frac{{V_{\rm oc}'}^2 R_{\rm load}}{(R_{\rm TEG}+R_{\rm load})^2},
\end{equation}

\noindent The maximization of the produced output power for a given thermal configuration therefore corresponds to:
\begin{equation}\label{adaptelec}
R_{\rm load}=R_{\rm TEG},
\end{equation}

\noindent which expressed in a more conventional way using the ratio $m = R_{\rm load}/R$ defined by Ioffe \cite{Ioffe}, reads:
\begin{equation}\label{rload}
\nonumber
m_{_{P=P_{\rm max}}} = 1 + \frac{Z\overline{T}}{K_{\rm contact}/K_{_{I=0}}+1},
\end{equation}

We see, as did Freunek and co-workers\cite{Freunek}, that the electrical impedance matching (\ref{rload}) does not correspond to the condition $m=1$ (or, equivalently, $R_{\rm load}=R$) of the ideal case since the equivalent resistance $R_{\rm TEG}$ of the generator has an additional part due to the finite thermal contact coupling. When the electrical resistance matching is satisfied, the maximum ouput power reads:
\begin{equation}\label{pmax}
P_{\rm max} = \frac{(K_{\rm contact}\Delta T)^2}{4(K_{_{I=0}}+K_{\rm contact})\overline{T}}~\frac{Z\overline{T}}{1+Z\overline{T}+K_{\rm contact}/K_{_{I=0}}},
\end{equation}

\subsection{Maximization of power by thermal impedance matching}

If we suppose that the TEG is used in a particular environment which imposes fixed thermal conductances for the contacts, we have to answer the question: how the thermal properties of the TEG can be chosen so that a maximum output power is obtained? This question directly relates to the general problem of the optimization of the working conditions of a non-endoreversible engine with heat exchangers coupled to the temperature reservoirs. This framework extends the classical so-called Novikov-Curzon-Ahlborn configuration \cite{Chambadal,Novikov,curzahl,bejan96} specialized to endoreversible engines, which implies that the heat exchangers are the only location for entropy production, a process thus governed by only one degree of freedom. In a non-endoreversible engine, entropy is produced inside the engine, and so in two different ways: the Joule effect and the thermal conduction effect; this confers two additional degrees of freedom to the system.

If we consider that $K_{\rm contact}$ is fixed by an external constraint, optimization of power may be achieved with respect to $K_{_{I=0}}$; calculations yield the condition:
\begin{equation}\label{thermaladapt}
\nonumber
\frac{K_{\rm contact}}{K_{_{I=0}}} = 1 + \frac{Z\overline{T}}{1+m},
\end{equation}

\noindent which corresponds to the equality: 
\begin{equation}\label{kctcteg}
K_{\rm contact} = K_{\rm TEG}
\end{equation}

\noindent Equation (\ref{kctcteg}) is similar to that derived by Stevens in Ref.~\cite{Stevens2001} where the thermal impedance matching corresponds to the equality between the thermal contact resistance and that of the TEG; however the difference with our result above is that the thermal resistance used in Ref.~\cite{Stevens2001} for the thermoelectric module is obtained under open circuit condition and, as such, it does not account for the convective part of the thermal current, while $K_{\rm TEG}$ defined in Eq.~(\ref{IQ2}) does. We end this part by highlighting the symmetry between electrical and thermal impedance matching respectively given by equations (\ref{rload}) and (\ref{thermaladapt}).

\subsection{Simultaneous thermal and electrical impedance matching}

The optimal point for a TEG offering the possibility to use two degrees of freedom: one electrical, $R_{\rm load}$, the other thermal, $K_{_{I=0}}$ (still with $Z\overline{T}$ fixed), in a particular configuration imposed by the environnement, is found by joint optimization of the electrical and thermal conditions. In other words, equations (\ref{rload}) and (\ref{thermaladapt}) have to be solved simultaneously. We find:
\begin{eqnarray}\label{k0ZT}
\frac{K_{\rm contact}}{K_{_{I=0}}} & = & \sqrt{Z\overline{T}+1}
\\
\label{mZT}
m_{_{P=P_{\rm max}}} & = & \sqrt{Z\overline{T}+1}
\end{eqnarray}

\noindent We note that Eq.~(\ref{k0ZT}) was put forward by Freunek and co-workers \cite{Freunek}, and that Yazawa and Shakouri obtained both equations \cite{Yasawa}. With these two impedance matching conditions, we find that the maximum power produced by the TEG is given by:
\begin{equation}
P_{\rm max} = \frac{K_{\rm contact} Z\overline{T}}{\left(1+\sqrt{1+Z\overline{T}}\right)^2} \frac{(\Delta T)^2}{4\overline{T}},
\end{equation}

\subsection{On the importance of thermal impedance matching}

\begin{figure}
\scalebox{.3}{\includegraphics*{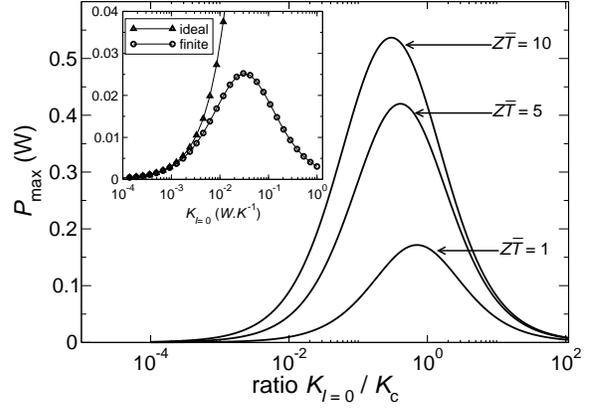}}
\caption{\label{fig2} Maximum power as function of the ratio $K_{_{I=0}}/K_{\rm contact}$ for various $Z\overline{T}$ values at fixed $K_{\rm contact}$. In the inset, the curves (with ideal and finite thermal contacts) are computed with the data of Ref.~\cite{nemirbeck} where the authors studied $P_{\rm max}$ for three values of $K_{_{I=0}} : 3\times 10^{-3},~ 6\times 10^{-3}$ and $1.2\times 10^{-2}$ W$\cdot$K$^{-1}$.}
\end{figure}

\noindent The variations of the maximum power $P_{\rm max}$ as a function of the ratio $K_{_{I=0}}/K_{\rm contact}$ [Eq.~(\ref{pmax})] are shown in Fig.~\ref{fig2} for three values of the figure of merit $Z\overline{T}$. For a given $Z\overline{T}$, $P_{\rm max}$ displays a bell-shape appearance (notice the use of a logarithmic scale for the abscissa axis). As can be expected, higher values of $Z\overline{T}$ yield greater values for the maximum of $P_{\rm max}$ and larger widths at half maximum. The maxima are shifted towards the region of lower values of the ratio $K_{_{I=0}}/K_{\rm contact}$. Figure~\ref{fig2} demonstrates the importance of thermal impedance matching: a high value of $Z\overline{T}$ does not guarantee a greater $P_{\rm max}$ for any value of the thermal conductance at zero electrical current $K_{_{I=0}}$: for instance, $P_{\rm max}$ at $K_{_{I=0}}=K_{\rm contact}$ for $ZT=1$ is greater than $P_{\rm max}$ at $K_{_{I=0}}= 5K_{\rm contact}$ for $ZT=10$. 

In the inset of Fig.~\ref{fig2}, two curves represent the maximum power as a function of $K_{_{I=0}}$ for \emph{finite} and perfect thermal contacts respectively; these shed light on the observations of Nemir and Beck \cite{nemirbeck}, which we mentionned in the Introduction: to analyze the impact of thermal contacts on device performance, they considered various configurations giving the same value for the figure of merit $Z{\overline T}$. They concluded that for a given value of contact thermal conductance the impact on the performance is strongly influenced by how the fixed figure of merit of the thermoelectric module is achieved. With our analysis one can now understand why the TEG with the highest $K_{_{I=0}}$ presents the largest performance degradation.

For fixed values of the electrical resistance and Seebeck coefficient of the module, electrical impedance matching implies that $Z\overline{T}$ is not constant, hence the optimization amounts to obtain the lowest possible value of the thermal conductivity inducing an increase of $Z\overline{T}$, which overcompensates the mismatching.

\subsection{Maximum efficiency}

The conversion of the heat current $I_{Q}$ into the electric power $P$ is a process characterized by the efficiency $\eta = P/I_{Q}$, which we write as:
\begin{equation}\label{etag}
\eta = \frac{K_{\rm contact}+K_{\rm TEG}}{K_{\rm contact}K_{\rm TEG}}\frac{P}{\Delta T},
\end{equation}

\noindent considering Eqs. (\ref{IQ2}) and (\ref{kapctc}). If the thermal contacts are ideal, then the expression above reduces to
\begin{equation}
\eta = \eta_{\rm C} \times \frac{m}{1 + m + \left( ZT_{\rm hot}\right)^{-1}\left( 1+ m \right)^2 - \eta_{\rm C}/2}
\end{equation}

We find that the value of $m$ (or equivalently the load resistance) which maximizes the efficiency (\ref{etag}) is:
\begin{equation}\label{metamax}
m_{_{\eta=\eta_{\rm max}}} = \sqrt{\left(1+Z\overline{T}\right)\left(1+Z\overline{T} \frac{K_{_{I=0}}}{K_{\rm contact}+K_{_{I=0}}}\right)}
\end{equation}

\noindent It explicitly depends on the thermal conductances  $K_{\rm contact}$ and $K_{_{I=0}}$. If the working conditions lead to modifications of $K_{\rm contact}$ (as, e.g., for liquid-gas heat exchangers), the operating point of the thermoelectric device changes consequently. It is thus interesting to see if $m_{_{\eta=\eta_{\rm max}}}$ is bounded when the ratio $K_{_{I=0}}/K_{\rm contact}$ varies. We checked from Eq.~(\ref{temps}) that the mean temperature $\overline{T}$ varies very little with $K_{\rm contact}$ so we may safely consider that the figure of merit is fixed without loss of generality for the discussion that follows.

\subsection{Analysis of optimization and power-efficiency trade-off}

\begin{figure}
\scalebox{.3}{\includegraphics*{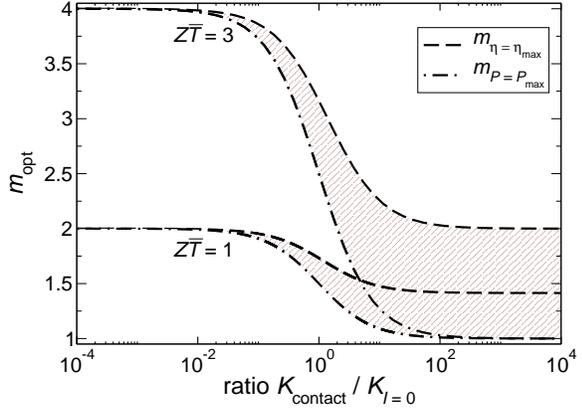}}
\caption{\label{fig3}Variations of the optimal parameters $m_{_{\eta=\eta_{\rm max}}}$ (dashed line) and $m_{_{P=P_{\rm max}}}$ (dashed-dotted line) as functions of $K_{\rm contact}$ scaled to $K_{_{I=0}}$, for $Z\overline{T}=1$ and $Z\overline{T}=3$. The shaded areas corresponds to the region of best optimization.}
\end{figure}

Accounting for finite thermal contacts in the TEG model induces changes in optimal values of the electrical load to achieve maximum power or efficiency. The optimal parameters $m_{\rm opt}$ [$m_{_{\eta=\eta_{\rm max}}}$ for maximum efficiency in Eq.~(\ref{metamax}), and $m_{_{P=P_{\rm max}}}$ for maximum power, in Eq.~(\ref{rload})] are plotted against $K_{\rm contact}$ (scaled to $K_{_{I=0}}$) in Fig. \ref{fig3} considering two values of the figure of merit: $Z\overline{T}=1$ and $Z\overline{T}=3$. For a given $Z\overline{T}$ and $K_{\rm contact} \gg K_{_{I=0}}$ (conditions close to perfect thermal contacts) the maximum power and maximum efficiency are well separated: $m_{_{\eta=\eta_{\rm max}}} \longrightarrow \sqrt{1+Z\overline{T}}$, and $m_{_{P=P_{\rm max}}} \longrightarrow 1$ (the separation between both obviously increases with $Z\overline{T}$).

Conversely, for $K_{\rm contact} \ll K_{_{I=0}}$ we obtain $m_{_{\eta=\eta_{\rm max}}} \longrightarrow 1+Z\overline{T}$, which also is the upper bound to $m_{_{P=P_{\rm max}}}$: both optimal parameters coincide. At first glance, the convergence of both optimal parameters towards the same value can be seen as valuable since this implies that there is no compromise to make between efficiency and power; however this is also results in a large performance decrease. Actually the regions lying between each pair of curves can be considered as the optimal regions to satisfy the power-efficiency trade-off. From this point of view we see that the narrowing of this zone, which also comes along with lower values of $Z\overline{T}$, is not at all desirable as it offers less flexibility in terms of working conditions of the thermoelectric generator.

\begin{figure}
\scalebox{.3}{\includegraphics*{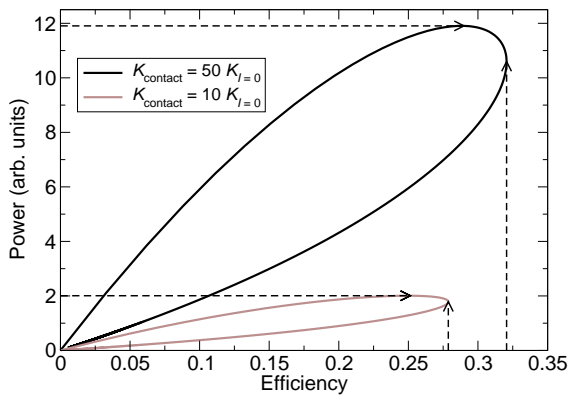}}
\caption{\label{fig4}Power versus effcicency curves for two cases with a fixed figure of merit $Z\overline{T} = 1$. Only the value of $K_{\rm contact}$ varies.}
\end{figure}

The figure \ref{fig4} displays two power-efficiency curves, one for $K_{\rm contact} = 10 K_{_{I=0}}$, the other for $K_{\rm contact} = 50 K_{_{I=0}}$. The narrowing of the optimal zone as the contact thermal conductance decreases is evidenced and thus confirms the observations made for Fig. \ref{fig3}. The arrows indicate the maximal values of $P_{\rm max}$ and $\eta_{\rm max}$ in the two cases. The maximum power is more sensitive than the maximum efficiency to the quality of the thermal contact: in the example we give, the ratio of the two highest values of $P_{\rm max}$ (one for each curve) is around 6, while the ratio of the two highest values of $\eta_{\rm max}$ is around 1.15. When the contact is not good enough the power-efficiency curve reduces to a point located at the origin where the efficiency and power optimisation are indeed identical.

\section{Discussion and conclusion}
Increasing the efficiency of thermoelectric conversion processes in real devices remains a topical research problem after several decades of efforts and progress. Minimization of entropy production methods have proved fruitful along the years in various sectors of thermal engineering and sciences \cite{bejan96} and new contributions are added to this field with, e.g., the recent introduction of the thermoelectric potential \cite{snyder2003,goupil1,goupil2}.

Thermoelectric efficiency has been theoretically shown to approach the Carnot efficiency in the case of an ergodic Lorentz gas with a large figure of merit \cite{casati}, but for real engines it is the efficiency at maximum power that is sought; such efficiency was initially shown to be bounded by the so-called formula of Curzon and Ahlborn in the specific case of an endoreversible heat engine \cite{curzahl} in the frame of finite-time themodynamics. The analysis of Curzon and Ahlborn was later put on firmer grounds in the frame of linear irreversible thermodynamics assuming a strong coupling between the heat flux and the work \cite{vandenbroeck}; applications to a nanoscopic quantum dot system \cite{esposito} and extension to stochastic heat engines \cite{schmiedl} followed. All these works assume strong or perfect couplings.

Using a force-flux formalism, we obtained the thermal and electrical conditions which allow the maximum power production by a thermoelectric generator \emph{non-ideally} coupled to heat reservoirs. Introducing an equivalent thermal conductance for the generator, which includes conductive and convective heat transports, we showed that the thermal impedance matching can be expressed in a simple fashion: the equality of the contact thermal conductances and the equivalent thermal conductance. Our analysis is thus physically more transparent than that based on the geometry of the elements that compose the thermoelectric generator.

Our calculations show that the interplay between the thermal and electrical properties of the TEG makes difficult the search for the optimum conditions for maximum output power production. The importance of the quality of the contacts between the TEG and the heat reservoirs is demonstrated: high values of $Z\overline{T}$ are of limited interest otherwise; in fact, one should search for the $Z\overline{T}$ that allows both electrical and thermal impedance matching, which is also the basic idea of the compatibility approach \cite{snyder2003} for ideal thermoelectric systems.

\acknowledgments
This work is part of the CERES2 and ISIS projects funded by the Agence Nationale de la Recherche. We thank Dr F. Mazzamuto and Dr W. Seifert for useful comments.


\begin{thebibliography}{0}

\bibitem{rowe}
  \Editor{Rowe D. M.}
  \Book{Thermoelectric Handbook, Macro to Nano}
  \Publ{CRC Press, Taylor and Francis Group}
  \Year{2006}.

\bibitem{disalvio}
  \Name{Di Salvio F. J.}
  \REVIEW{Science}{285}{1999}{703}.
  
\bibitem{shakouri}
  \Name{Shakouri A.}
  \REVIEW{Annu. Rev. Mater. Res.}{41}{2011}{399}.

\bibitem{snyder1}
  \Name{Pei Y., Shi X., LaLonde A., Wang H., Cheng L. \and Snyder G. J.}
  \REVIEW{Nature}{473}{2011}{66}.

\bibitem{nemirbeck} 
  \Name {Nemir D. \and Beck J.} 
  \REVIEW{J. Electon. Mat.}{39}{2010}{1897}.

\bibitem{Chambadal}
  \Name{Chambadal P.}
  \Book{Les centrales nucl\'eaires}
  \Publ{Armand Colin}
  \Year{1957}.

\bibitem{Novikov}
  \Name{Novikov I. I.}
  \REVIEW{J. Nucl. En.}{7}{1958}{125}.

\bibitem{curzahl}
  \Name{Curzon F. \and Ahlborn B.}
  \REVIEW{Am. J. Phys}{43}{1975}{22}.

\bibitem{bejan96}
  \Name{Bejan A.}
  \REVIEW{J. Appl. Phys.}{79}{1996}{1191}.

\bibitem{clingman}
  \Name{Clingman W.}
  \REVIEW{Adv. En. Conv.}{1}{1961}{61}.

\bibitem{snyder04}
  \Name{Snyder G. J.}
  \REVIEW{Appl. Phys. Lett.}{84}{2004}{2436}.

\bibitem{seifert10}
  \Name{Seifert W., Zabrocki K., M\"uller E.  \and Snyder G. J.}
  \REVIEW{phys. stat. sol. (a)}{207}{2010}{2399}.

\bibitem{onsager1}
  \Name{Onsager L.}
  \REVIEW{Phys. Rev.}{37}{1931}{405}.

\bibitem{onsager2}
  \Name{Onsager L.}
  \REVIEW{Phys. Rev.}{38}{1931}{2265}.

\bibitem{callen1}
  \Name{Callen H. B.}
  \REVIEW{Phys. Rev.}{73}{1948}{1349}.
  
\bibitem{callen2}
  \Name{Callen H. B.}
  \REVIEW{Rev. Mod. Phys.}{26}{1954}{237}.

\bibitem{huleihil}
  \Name{Huleihil M. \and Andresen B.}
  \REVIEW{J. Appl. Phys.}{100}{2006}{014911}.

\bibitem{snyder2003}
  \Name{Snyder G. J. \and Ursell T. S.}
  \REVIEW{Phys. Rev. Lett.}{91}{2003}{148301}.

\bibitem{goupil2}
  \Name{Goupil C., Seifert W., Zabrocki K., M\"uller E. \and Snyder G. J.}
  \REVIEW{Entropy}{13}{2011}{1481}.

\bibitem{Ioffe}
  \Name{Ioffe A.}
  \Book{Semiconductor thermoelements and thermoelectric cooling}
  \Publ{London, Infosearch, ltd.}
  \Year{1957}.
  
\bibitem{Freunek}
  \Name{Freunek M., M\"uller M., Ungan T., Walker W. \and Reindl L. M.}
  \REVIEW{J. Electron. Mat.}{38}{2009}{1214}.

\bibitem{Yasawa}
  \Name{Yasawa K. \and Shakouri A.}
  \REVIEW{Environ. Sci. Technol.}{45}{2011}{7548}.
  
\bibitem{Stevens2001}
  \Name{Stevens J. W.}
  \REVIEW{En. Conv. Manag.}{42}{2001}{709}.

\bibitem{goupil1}
  \Name{Goupil C.}
  \REVIEW{J. Appl. Phys.}{106}{2009}{104907}.

\bibitem{casati}
  \Name{Casati G., Mej\'ia-Monasterio C. \and Prosen T.}
  \REVIEW{Phys. Rev. Lett.}{101}{2008}{016601}.

\bibitem{vandenbroeck}
  \Name{Van den Broeck C.}
  \REVIEW{Phys. Rev. Lett.}{95}{2005}{190602}.

\bibitem{esposito}
  \Name{Esposito M., Lindenberg K. \and Van den Broeck C.}
  \REVIEW{Europhys. Lett.}{85}{2009}{60010}.

\bibitem{schmiedl}
  \Name{Schmiedl T. \and Seifert U.}
  \REVIEW{Europhys. Lett.}{81}{2008}{20003}.

\end{thebibliography}
\end{document}